\documentclass[twocolumn,prb,showpacs,showkeys,amsmath,amssymb,floatfix]{revtex4}

\usepackage{graphicx} % Include figure files
\usepackage{dcolumn}  % Align table columns on decimal point
\usepackage{bm}       % bold math
\usepackage{natbib}

\usepackage{psfrag}   % psfrag
\usepackage{pifont}
\usepackage{gensymb,wasysym}
\usepackage{tensind}  % for tensor notation

\def\arsinh{\text{arsinh}}

\def\artanh{\text{artanh}}
\def\abs#1{|#1|}
\def\sign{\text{sign}}

\def\thms#1#2#3{{#1}^{\text{h}}{#2}^{\text{min}}{#3}^{\text{sec}}}
\def\tj#1{{#1}^{\text{a}}}

\def\jApp{\ding{43}{\it Java-Applet}}

\begin{document}

\preprint{unknown}
\title{A trip to the end of the universe and the twin ``paradox''}

\author{Thomas M{\"u}ller, Andreas King and Daria Adis}
%\homepage{http://www.tat.physik.uni-tuebingen.de/~tmueller}
\email{tmueller@tat.physik.uni-tuebingen.de}
\affiliation{
  Institut f{\"u}r Astronomie und Astrophysik, 
  Abteilung Theoretische Astrophysik\\
  Auf der Morgenstelle 10,
  72076 T{\"u}bingen, Germany
}
% -----------------------------------------------------------------
%                            Abstract
% -----------------------------------------------------------------

\begin{abstract}
  In principle, the twin paradox offers the possibility to go on a trip to the center of our galaxy or even to the end of our universe within life time. In order to be a most comfortable journey the voyaging twin accelerates with Earth's gravity. We developed some Java applets to visualize what both twins could really measure, namely time signals and light coming from the surrounding sky.
\end{abstract}

% 03.30.+p   Special relativity
% 07.05.Rm   Data presentation and visualization: algorithms and implementation

\maketitle

% -----------------------------------------------------------------
%                            Introduction
% -----------------------------------------------------------------
\section{\label{sec:level1}Introduction}
We take the view that the twin paradox is well understood and we won't loose time explaining it once more. For a detailed discussion of this topic we refer the reader to standard literature. The crucial detail in the twin paradox is the fact that there is an asymmetry between both twins. Accepting this makes the word ``paradox'' meaningless.\par
It is the purpose of this article to emphasize the local perspective of each twin who could only do measurements with respect to their individual reference frames. This resembles Bondi's k-calculus\cite{dinverno} or the concept of radar time as discussed by Dolby and Gull\cite{dolby2001} where hypersurfaces of simultaneity are defined by local measurements of reflected light signals. The interesting information both twins can interchange in our example are their individual proper time. But they will not be able to determine the time dilation out of that because the information carrying the proper time of the other twin needs some time to travel the distance between them. This time interchange, as the motion itself, is not symmetric. Besides the time signals, both twins also receive light from distant stars or other astronomical sources. Depending on the relative velocity with respect to the light source, each twin will have a different view of the stellar sky because of aberration and Doppler shift.\par
It is a well known fact that acceleration is not the crucial basis to explain the twin paradox. Only the difference in length of their paths make the twins age at different rates. But for a complete journey the traveling twin has to leave Earth with a rocket starting with zero velocity. Having reached his destination the twin might return to Earth --- if Earth still exists. In order to be a most comfortable journey we consider a uniformly accelerated motion which is separated into four phases of equal period. Even though acceleration is usually brought into connection with general relativity\cite{perrin1979,good1982} it can also be treated only in special relativity\cite{french}.\par
We expect the reader to be familiar with basic calculations like the Lorentz transformations and the description of acceleration within special relativity.
In Sec.~\ref{sec:unifAcc} we review the description of uniformly accelerated motion in special relativity. Following the example of Ruder\cite{ruder} we present a special round trip consisting of four equal acceleration phases in Sec.~\ref{sec:roundTrip}. From a local perspective taken in Sec.~\ref{sec:timeSignals} only time or light signals can be measured by each twin. As an example, we consider a flight to Vega and examine the time signal exchange in Sec.~\ref{sec:flightVega}. In Sec.~\ref{sec:accFrame} we derive the frequency shift and aberration formulas for the following visualization of the stellar sky. In Sec.~\ref{sec:visSky} we describe the visual effects due to aberration and frequency shift. In the last section we show how far we can travel with the help of time dilation and length contraction.
% -----------------------------------------------------------------
%                           uniform acceleration
% -----------------------------------------------------------------
\section{\label{sec:unifAcc}Uniform acceleration}
The earth twin Eric stays in the inertial reference frame $S$, whereas his traveling twin sister Tina leaves the Earth with constant acceleration $\alpha$ with respect to her own instantaneous system $S'$. Please note that all primed quantities refer to $S'$, and unprimed quantities to $S$. From the Newtonian point of view, Tina will reach a velocity $v=\alpha\Delta t$ within the time $\Delta t$. She will then have covered a distance $\Delta x=\frac{1}{2}\alpha\Delta t^2$. Both twins will agree on time, velocity and distance. For velocities much smaller than the speed of light, $v\ll c$, this will be still correct. But within special relativity the earth twin Eric would measure an acceleration \cite{french,rindler}
\begin{equation}
  \label{eq:srtAccel}
  a = \alpha\left(1-\beta^2\right)^{3/2} = \frac{\alpha}{\gamma(\beta)^3}
\end{equation}
with $\beta=v/c$ and $\gamma=1/\sqrt{1-\beta^2}$. Starting with velocity $v_0$, we get Tina's velocity $v$ at time $t$ by substituting $a=dv/dt$ in Eq.(\ref{eq:srtAccel}). Integration results in
\begin{equation}
  \label{eq:srtVel}
  v = \frac{\alpha t+c\zeta}{\sqrt{1+\left(\alpha t/c+\zeta\right)^2}}.
\end{equation}
with $\zeta=\gamma(\beta_0)\beta_0$. The traveled distance $x=x(t)$ follows from Eq.(\ref{eq:srtVel}) by another integration,
\begin{equation}
  \label{eq:srtDist}
  x = \frac{c^2}{\alpha}\left[\sqrt{1+\left(\frac{\alpha t}{c}+\zeta\right)^2}-\sqrt{1+\zeta^2}\right] + x_0,
\end{equation}
where $x_0$ is the starting position at time $t=0$.\par
Time in both systems will go by at different rates depending on the relative velocity $\beta$,
\begin{equation}
  dt'=\gamma(\beta)^{-1}dt.
\end{equation}
Thus, synchronizing $t=t'=0$ gives
\begin{equation}
  \label{eq:srtProperTime}
  t' = \frac{c}{\alpha}\left[\arsinh\left(\frac{\alpha t}{c}+\zeta\right)-\arsinh\zeta\right]
\end{equation}
or
\begin{equation}
  \label{eq:srtCoordTime}
  t = \frac{c}{\alpha}\left[\sinh\left(\frac{\alpha t'}{c}+\arsinh\zeta\right)-\zeta\right].
\end{equation}
Eq.(\ref{eq:srtCoordTime}) simplifies for $\beta_0=0$ to
\begin{equation}
  \label{eq:relTime}
  t = \frac{c}{\alpha}\sinh\frac{\alpha t'}{c}
\end{equation}
and if the traveling twin starts at $x_0=0$ its current position $x=x\left(t'\right)$ is given by
\begin{equation}
  \label{eq:currPos0}
  x = \frac{c^2}{\alpha}\left(\cosh\frac{\alpha t'}{c}-1\right).
\end{equation}
Tina's velocity, with respect to her proper time $t'$, follows from Eq.(\ref{eq:relTime}) together with Eq.~(\ref{eq:srtVel}),
\begin{equation}
  \label{eq:relVel}
  v = c\,\tanh\frac{\alpha t'}{c}.
\end{equation}
Hence, since $\abs{\tanh(x)}\leq 1$, locally measured speed is always less than the speed of light in accordance with special relativity.
% -----------------------------------------------------------------
%                      A specific round trip
% -----------------------------------------------------------------
\section{\label{sec:roundTrip}A specific round trip}
\begin{figure}[ht!]
  \begin{center}
     \includegraphics[scale=0.8]{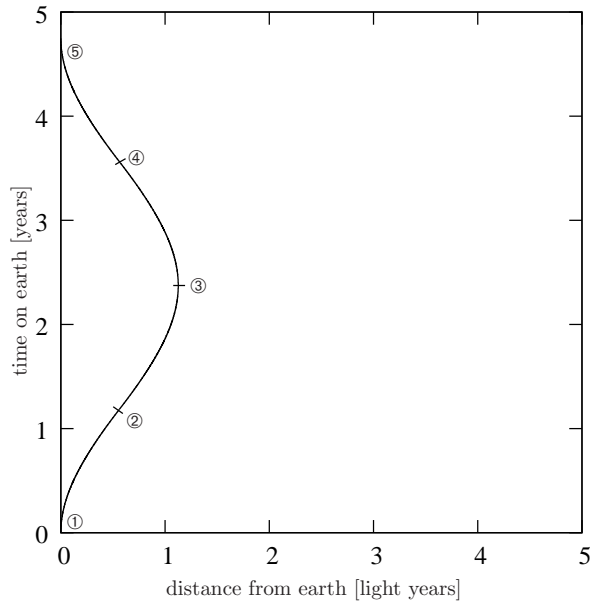}
     \caption{\label{fig:phases}Tina's journey is separated into four phases. She starts from point \ding{192}, accelerates up to maximum velocity at point \ding{193} and slows down until she reaches the turning point \ding{194}. Then she accelerates into the opposite direction and slows down again until she comes home.}
  \end{center}
\end{figure}

The situation we want to examine is the following. While Eric stays at home, Tina goes on a journey that is separated into four phases which last the time $T'$ each. In the first phase she starts at point \ding{192} and accelerates with $\alpha$ until she reaches the maximum velocity
\begin{equation}
  v_{\text{max}}=c\,\tanh\frac{\alpha T'}{c}
\end{equation}
at point \ding{193}, compare Fig.~\ref{fig:phases}. Next, she decelerates with $-\alpha$ until she stops at her destination \ding{194} where she has covered the distance
\begin{equation}
   x_{\text{max}} = 2\frac{c^2}{\alpha}\left(\cosh\frac{\alpha T'}{c}-1\right).
\end{equation}
The same procedure, but now in the opposite direction, brings Tina back to home. Thus, the hole trip takes $4T'$ with respect to Tina's proper time, whereas Eric has to wait the time
\begin{equation}
  \label{eq:completeTime}
  4T = 4\frac{c}{\alpha}\sinh\frac{\alpha T'}{c}
\end{equation}
for his sister's return. Hence, the bigger the acceleration is and the longer the journey lasts, the bigger is the effect of time dilation and the different aging of both twins. Tina's worldline is shown in a spacetime diagram (Fig.~\ref{fig:phases}), where the proper time $ct$ of Eric is plotted over Tina's distance $x$ to Eric. 

% -----------------------------------------------------------------
%                      Time signals
% -----------------------------------------------------------------
\section{\label{sec:timeSignals}Observation of time signals}
In contrast to normal discussions of special relativity concerning length contraction or time dilation with a group of synchronized observers, we concentrate on a local perspective. Hence, we can only use light signals exchanged by both twins to determine the proper time in each case. For this, the signals are coded with the proper emission time $t_{\text{emit}}$ or $t'_{\text{emit}}$ which will be received by the other twin at observation time $t'_{\text{obs}}$ or $t_{\text{obs}}$. Of course, this measured time is not the actual time of the other twin. Because of the finite speed of light the signals need some time to travel the distance between both twins.\par
\begin{figure}[ht]
  \begin{center}
    \includegraphics[scale=0.8]{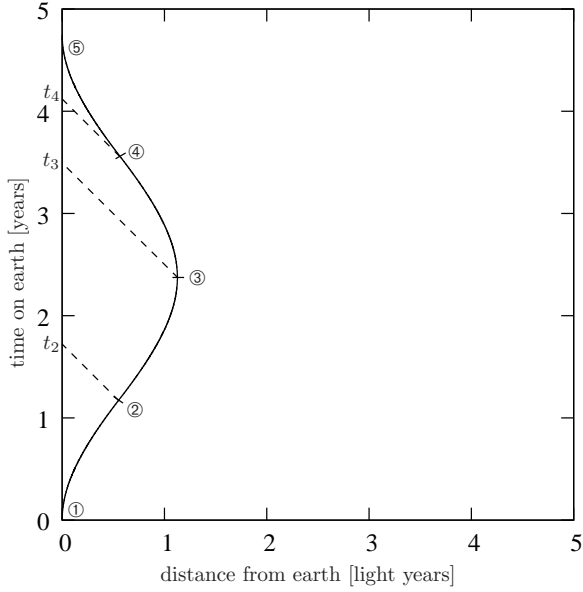}
    \caption{\label{fig:signals}Signals which were emitted by Tina in the accelerating period $i$ reach the earth twin Eric in the interval $[t_i,t_{i+1}]$.}
  \end{center}
\end{figure}

{\it Eric's perspective:} Eric observes at his proper time $t_{\text{obs}}$ a signal send from Tina. Because of the finite speed of light the signal must have been emitted at earth time $t_{\text{emit}}$,
\begin{equation}
  t_{\text{obs}} = t_{\text{emit}} + \frac{x\left(t_{\text{emit}}\right)}{c}.
\end{equation}
The difference between $t_{\text{obs}}$ and $t_{\text{emit}}$ is exactly the time needed for the signal to travel from Tina's current position $x_{\text{emit}}=x\left(t_{\text{emit}}\right)$ to the Earth. In order to determine the exact position $x_{\text{emit}}$ we have to find out the period of acceleration when Tina has emitted the signal (compare Fig.~\ref{fig:signals}).

The border points $t_i, (i=2,\ldots,4)$, of the intervals follow from conditions like $t_2 = t_{\text{\ding{193}}}+x_{\text{\ding{193}}}/c$, whereas $t_5$ follows from Eq.~(\ref{eq:completeTime}). Thus, we have
\begin{subequations}
  \begin{align}
    t_1 &= 0,\\
    t_2 &= \frac{c}{\alpha}\left(\sinh\frac{\alpha T'}{c}+\cosh\frac{\alpha T'}{c}-1\right),\\
    t_3 &= \frac{c}{\alpha}\left(2\sinh\frac{\alpha T'}{c}+2\cosh\frac{\alpha T'}{c}-2\right),\\
    t_4 &= \frac{c}{\alpha}\left(3\sinh\frac{\alpha T'}{c}+\cosh\frac{\alpha T'}{c}-1\right),\\
    t_5 &= 4\frac{c}{\alpha}\sinh\frac{\alpha T'}{c}.
  \end{align}
\end{subequations}
Depending on the interval $[t_i,t_{i+1}]$ Eric receives signals from the accelerating period $i$, where Tina's current position was $x_{\text{emit}}=x\left(t_{\text{emit}}\right)$. In the first period \ding{192}-\ding{193} we get from Eq.(\ref{eq:srtDist}) with $x_0=0$,
\begin{align}
\nonumber   t_{\text{obs}} &= t_{\text{emit}}+\frac{x_{\text{\ding{192}-\ding{193}}}\left(t_{\text{emit}}\right)}{c}\\
  \label{eq:tbfirst}
  &= t_{\text{emit}}+\frac{c}{\alpha}\left(\sqrt{1+\frac{\alpha^2t_{\text{emit}}^2}{c^2}}-1\right).
\end{align}
Solving Eq.(\ref{eq:tbfirst}) for $t_{\text{emit}}$ and putting it into Eq.(\ref{eq:srtProperTime}) with $\beta_0=0$ gives the observed signal time $t'_{\text{emit}}$,
\begin{equation}
  t'_{\text{emit}} = \frac{c}{\alpha}\arsinh\frac{\alpha t_{\text{emit}}}{c} = \frac{c}{\alpha}\ln\left(1+\frac{\alpha t_{\text{obs}}}{c}\right).
\end{equation}

In the accelerating period \ding{193}-\ding{195}, Eric's observation time $t_{\text{obs}}$ is related to the emission time $t_{\text{emit}}$ via
\begin{equation}
  \label{eq:tbsecond}
  t_{\text{obs}} = T+\tilde{t}_{\text{emit}}+\frac{x_{\text{\ding{193}-\ding{195}}}\left(\tilde{t}_{\text{emit}}\right)}{c},
\end{equation}
where $\tilde{t}_{\text{emit}}=t_{\text{emit}}-T$ starts measuring time from point \ding{193}. Tina's current position $x_{\text{\ding{193}-\ding{195}}}\left(\tilde{t}_{\text{emit}}\right)$ follows from Eq.(\ref{eq:srtDist}),
\begin{align}
  \nonumber  x_{\text{\ding{193}-\ding{195}}}\left(\tilde{t}_{\text{emit}}\right) = &-\frac{c^2}{\alpha}\left(\sqrt{1+\left(\frac{-\alpha\tilde{t}_{\text{emit}}}{c}+\zeta\right)^2}-\sqrt{1+\zeta^2}\right)\\
  &+\frac{c^2}{\alpha}\left(\sqrt{1+\frac{\alpha^2T^2}{c^2}}-1\right)
\end{align}
where $\zeta=\beta_0\gamma\left(\beta_0\right)$ and the velocity $\beta_0$ is given by
\begin{equation}
  \beta_0 = \frac{\alpha T/c}{\sqrt{1+\alpha^2T^2/c^2}}.
\end{equation}
From Eq.(\ref{eq:tbsecond}) we can deduce $\tilde{t}_{\text{emit}}$ to be
\begin{equation}
  \tilde{t}_{\text{emit}} = \frac{1}{2}\frac{2\xi\sqrt{1+\zeta^2}-\xi^2\alpha/c}{\sqrt{1+\zeta^2}+\zeta-\xi\alpha/c}
\end{equation}
with the abbreviation
\begin{equation}
  \xi = t_{\text{obs}}-T-\frac{c}{\alpha}\left(\sqrt{1+\frac{\alpha^2T^2}{c^2}}-1\right)
\end{equation}
and Eq.(\ref{eq:srtProperTime}) gives the corresponding time $\tilde{t}'_{\text{emit}}$. Hence, Eric receives at his observation time $t_{\text{obs}}$ the emission time $t'_{\text{emit}}=\tilde{t}'_{\text{emit}}+T'$.\par
In the last period \ding{195}-\ding{196} Eric's observation time $t_{\text{obs}}$ follows from
\begin{equation}
  t_{\text{obs}} = \bar{t}_{\text{emit}}+3T+\frac{x_{\text{\ding{195}-\ding{196}}}\left(\bar{t}_{\text{emit}}\right)}{c},
\end{equation}
where $x_{\text{\ding{195}-\ding{196}}}\left(\bar{t}_{\text{emit}}\right)$ equals $x_{\text{\ding{193}-\ding{195}}}\left(\tilde{t}_{\text{emit}}\right)$ with substitutions $\alpha\mapsto-\alpha, \zeta\mapsto -\zeta$ and $\tilde{t}_{\text{emit}}\mapsto\bar{t}_{\text{emit}}$ and $\bar{t}_{\text{emit}}=t_{\text{emit}}-3T$. In order to get the emission time $t_{\text{emit}}$ out of $t_{\text{obs}}$ we have to follow the same procedure like in the previous accelerating period.\par
{\it Tina's perspective:} Now, we consider the opposite situation where Eric sends his proper time $t_{\text{emit}}$ to Tina. The arrival time $t'_{\text{obs}}$ of the signal just follows from intersecting the future light cone of Eric at time $t_{\text{emit}}$ with Tina's world line,
\begin{equation}
  \label{eq:obsLightCone}
  t_{\text{obs}}=t_{\text{emit}}+\frac{x_{\text{obs}}}{c}.
\end{equation}
In contrast to Eric's perspective the calculations are a little bit more straight forward because we determine $t_{\text{emit}}$ out of $t'_{\text{obs}}$. While Tina is in the first accelerating phase \ding{192}-\ding{193} her proper time $t'_{\text{obs}}$ transforms into $t_{\text{obs}}$ via Eq.(\ref{eq:relTime}). Thus, her position is given by
\begin{equation*}
  x_{\text{obs}} = \frac{c^2}{\alpha}\left(\cosh\frac{\alpha t'_{\text{obs}}}{c}-1\right)
\end{equation*}
and the signal from Eric which arrives at time $t'_{\text{obs}}$ has the time signature 
\begin{equation*}
  t_{\text{emit}} = \frac{c}{\alpha}\left(\sinh\frac{\alpha t'_{\text{obs}}}{c}-\cosh\frac{\alpha t'_{\text{obs}}}{c}+1\right).
\end{equation*}

In the accelerating phase \ding{193}-\ding{195} we get $\tilde{t}_{\text{obs}}=t_{\text{obs}}-T$ from Eq.~(\ref{eq:srtCoordTime}) with $\zeta=\sinh\left(\alpha T'/c\right)$ for the observation time $\tilde{t}'_{\text{obs}}=t'_{\text{obs}}-T'$. Thus, we have
\begin{equation*}
  t_{\text{obs}} = T -\frac{c}{\alpha}\left(\sinh\frac{\alpha \left(-t'_{\text{obs}}+2T'\right)}{c}-\sinh\frac{\alpha T'}{c}\right).
\end{equation*}
Tina's position $x_{\text{obs}}$, when she receives the signal, is given by Eq.~(\ref{eq:srtDist}) which simplifies to
\begin{equation*}
  x_{\text{obs}}=-\frac{c^2}{\alpha}\left(\cosh\frac{\alpha\left(-t'_{\text{obs}}+2T'\right)}{c}-2\cosh\frac{\alpha T'}{c}+1\right).
\end{equation*}
The same procedure also applies to the last accelerating phase \ding{195}-\ding{196} where we get
\begin{equation*}
  t_{\text{obs}} = 3T+\frac{c}{\alpha}\left[\sinh\frac{\alpha\left(t'_{\text{obs}}-4T'\right)}{c}+\sinh\frac{\alpha T'}{c}\right]
\end{equation*}
for the observation time and
\begin{equation*}
  x_{\text{obs}} = \frac{c^2}{\alpha}\left[\cosh\frac{\alpha\left(t'_{\text{obs}}-4T'\right)}{c}-1\right]
\end{equation*}
for Tina's position. In both cases, the emission time $t_{\text{emit}}$ follows from Eq.~(\ref{eq:obsLightCone}) with the corresponding time $t_{\text{obs}}$ and position $x_{\text{obs}}$.
% -----------------------------------------------------------------
%           Flight to Vega
% -----------------------------------------------------------------
\section{\label{sec:flightVega}Flight to Vega}
\begin{figure}[ht!]
  \begin{center}
     \includegraphics[scale=0.8]{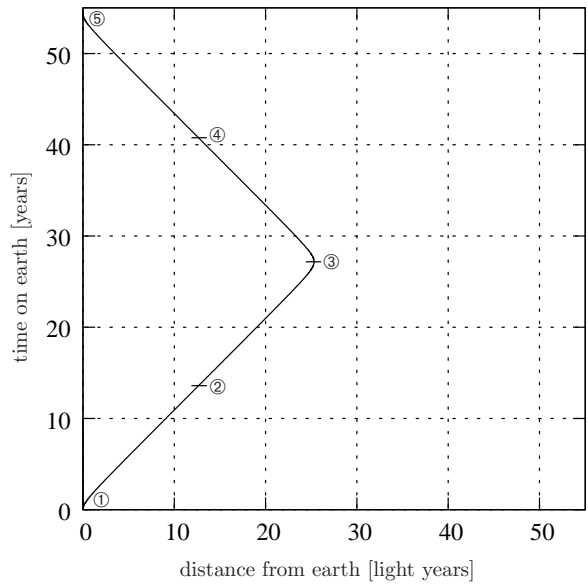}
     \caption{\label{fig:stDia}Spacetime diagram for Tina's flight to Vega and return to Earth with respect to Eric's frame. At event \ding{194} Tina reaches Vega and immediately returns. At \ding{193} and \ding{195} she changes her acceleration direction.}
  \end{center}
\end{figure}

As a first example we consider a flight to our neighboring star Vega, which is roughly $25.3~ly$ away\footnote{One light year $(1~ly)$ equals the distance which is covered by light with velocity $c=299,792,458~m/s$ within one year.}. In order to be a most comfortable journey Tina accelerates with $\alpha=1g\approx 9.81~m/s^2$. While the journey takes $T'_{\text{total}}=4T'\approx 12.93$ years with respect to her proper time, Eric has to wait $T_{\text{total}}\approx 54.48$ years for his sister. The worldline of Tina with respect to Eric's rest frame is shown in the spacetime diagram, Fig.~\ref{fig:stDia}. Even though Tina follows an accelerated motion, her worldline looks to be almost linear. 

\begin{figure}[ht]
  \begin{center}
    \includegraphics[scale=0.73]{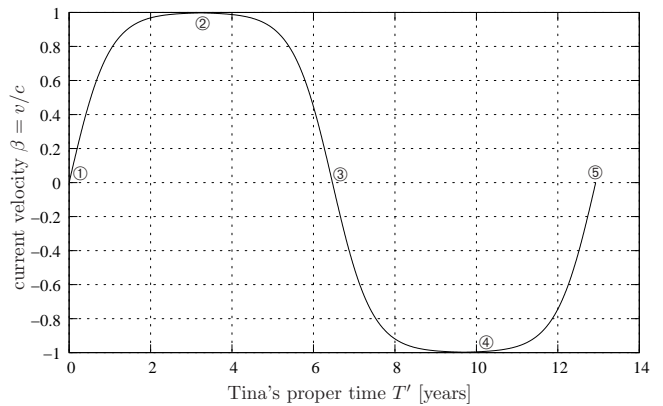}
    \caption{\label{fig:currVel}Tina's current velocity with respect to her own proper time. Maximum speed $\abs{\beta}\approx 0.9975$ is reached in Point \ding{193} and Point \ding{195}.}
  \end{center}
\end{figure}

Because of the moderate acceleration Tina's rocket needs roughly one year to get $80\%$ of the speed of light. At position $x_{\text{\ding{193}}}$ she reaches maximum speed $\beta_{\text{max}}\approx 0.9975$, compare Fig.~\ref{fig:currVel}.

\begin{figure}[ht]
  \begin{center}
    \includegraphics[scale=0.83]{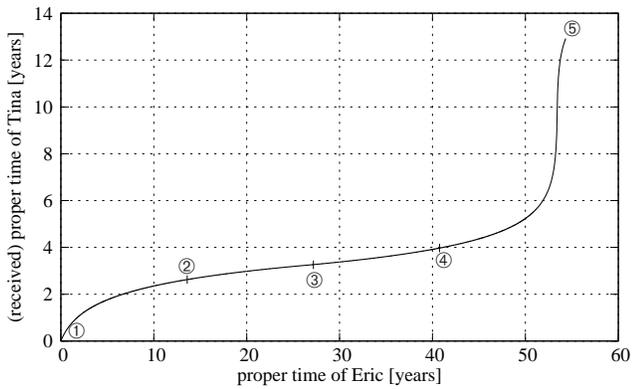}
    \caption{\label{fig:tsReceived}The earth twin Eric sees/receives at his proper time $t_{\text{obs}}$ (abscissa) Tina's proper time $t'_{\text{emit}}$ (ordinate).}
  \end{center}
\end{figure}

\begin{figure}[ht]
  \begin{center}
    \includegraphics[scale=0.95]{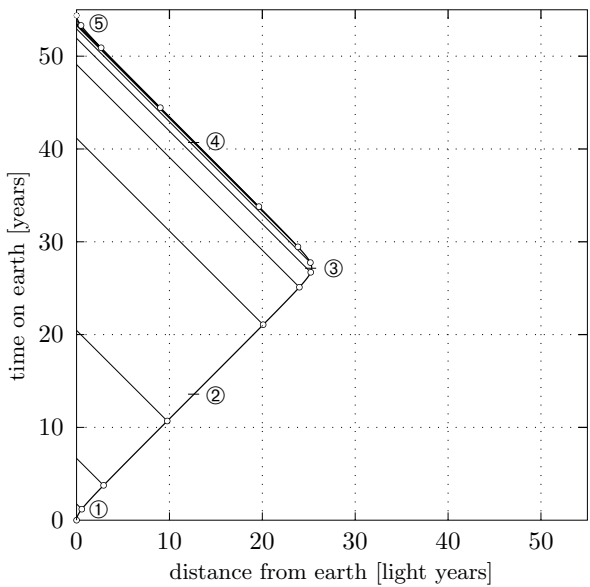}	
    \caption{\label{fig:tsSendSignal}The rocket twin Tina sends a time signal every year (small circles) with respect to her proper time $t'$. Note that time units are not equally spaced on Tina's worldline.}
  \end{center}
\end{figure}

\begin{figure}[ht]
  \begin{center}
    \includegraphics[scale=0.83]{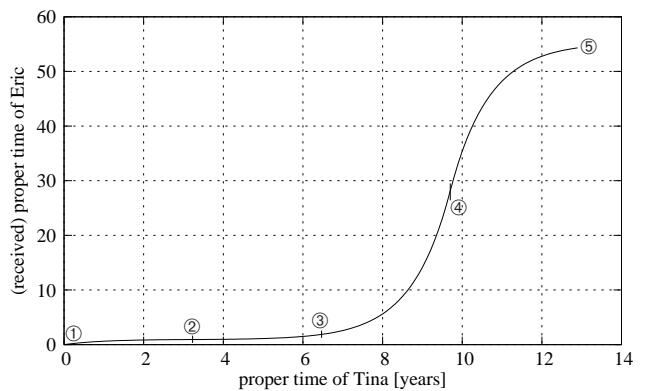}
    \caption{\label{fig:tReceived}The rocket twin Tina sees/receives at her proper time $t'_{\text{obs}}$ (abscissa) Eric's proper time $t_{\text{emit}}$ (ordinate).}
  \end{center}
\end{figure}

\begin{figure}[ht]
  \begin{center}
    \includegraphics[scale=0.95]{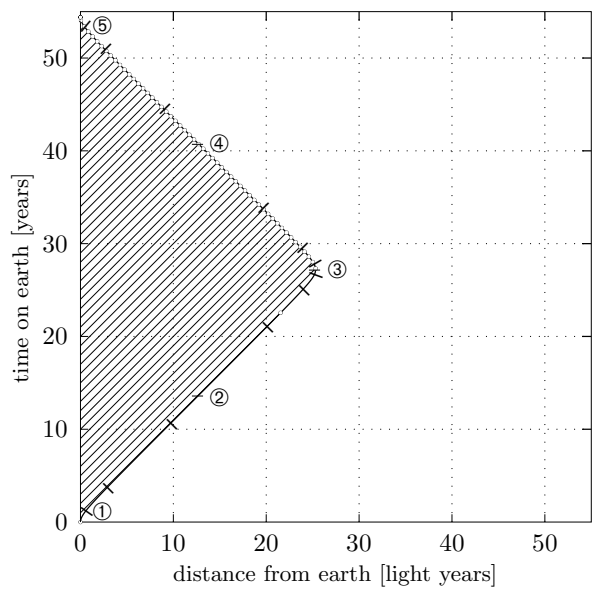}
    \caption{\label{fig:tSendSignal}The earth twin Eric sends a time signal every year with respect to his proper time $t$. The dashes on Tina's worldline mark the years of her proper time. Eric's first signal reaches Tina not until she already decelerates to her destination \ding{194}.}
  \end{center}
\end{figure}

Now, let's consider the exchange of time signals between both twins. Proper time sent by the one twin and received by the other is plotted in a time-time-diagram, compare Fig.~\ref{fig:tsReceived} and Fig.~\ref{fig:tReceived}. The proper time of the twin who receives a signal is plotted on the abscissa whereas the proper time, which was sent by the other twin, is shown on the ordinate. The shape of the curves are hard to understand because two effects are mixed, time dilation and nonlinear change of distance. In order to explain the differences between these curves, we also plotted the space-time diagram with Tina's worldline. The time signals are represented by straight lines with $45\degree$ slope, compare Fig.~\ref{fig:tsSendSignal} and Fig.~\ref{fig:tSendSignal}. Note that time intervals along Tina's worldline are not equally spaced because of her accelerated motion.\par
{\it Eric's perspective (Fig.~\ref{fig:tsReceived} and \ref{fig:tsSendSignal}):} While Eric stays at home, Tina leaves Earth with proper acceleration $\alpha$. Because the distance between both twins grows, it's quite obvious that signals emitted by Tina need more and more time to reach Eric. Additionally, because of time dilation, Tina seems to wait a longer time than expected before emitting the next signal. Even though Tina is already on the way back to home when she has left her destination \ding{194}, Eric will collect most of her time signals only in the very end of her journey.

{\it Example:} Eric observes at his proper time $t_{\text{obs}}\approx 20$ years a time signal which was sent by Tina at her proper time $t'_{\text{emit}}=3$ years when she was $10~ly$ away from home. But her current position at Eric's observation time is $x\approx 19~ly$ and her watch shows roughly $4$ years.

{\it Tina's perspective (Fig.~\ref{fig:tReceived} and \ref{fig:tSendSignal}):} It might be obvious that Tina's perspective is a little bit different. Eric's ``first year'' signal reaches Tina not until she already decelerates to her destination \ding{194}. On her way home, Tina seems to receive the signals quite regularly but because of time dilation she collects most of the signals around point \ding{195} when she has maximum speed and the time dilation effect is strongest.

In order to investigate this situation for different proper accelerations $\alpha$ and different periods of time $T'$, we have written an interactive Java applet (App.~\ref{app:javaApplet}). \jApp
% -----------------------------------------------------------------
%             The accelerated reference frame        
% -----------------------------------------------------------------
\section{\label{sec:accFrame}The accelerated reference frame}
%% ----------------------------------
%%        subsec: Aberration, Doppler shift and length contraction
%% ----------------------------------
\subsection{Aberration, Doppler shift and length contraction}
In four-dimensional spacetime any observer has their own local reference system which is given by four base vectors. Each measurement is taken with respect to this local tetrad.\par
\begin{figure}[ht]
  \begin{center}
    \includegraphics[scale=1.0]{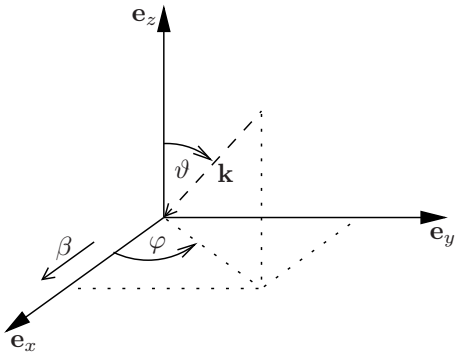}
    \caption{\label{fig:waveVector}Wave vector $\mathbf{k}$ of an incoming light ray in spherical coordinates $\vartheta\in(0,\pi), \varphi\in(-\pi,\pi)$ with respect to Eric's rest frame. The voyaging twin Tina is currently moving with velocity $\beta$ along the $\mathbf{e}_x$ direction.}
  \end{center}
\end{figure}

Consider the flat Minkowski spacetime which is represented by the metric
\begin{equation}
  \label{eq:minkowski}
  ds^2 = -c^2dt^2 + dx^2 + dy^2 + dz^2.
\end{equation}
The local tetrad $\left\{\mathbf{e}_i\right\}_{(i=0,\ldots,3)}$ of an observer moving with velocity $\beta=v/c$ in the direction of the $x$-axes reads
\begin{subequations}
\begin{alignat}{3}
  \mathbf{e}_0  &= \gamma\left(\mathbf{e}_t + \beta\mathbf{e}_x\right), &\qquad \mathbf{e}_2 &= \mathbf{e}_y,\\
   \mathbf{e}_1 &=\gamma\left(\beta\mathbf{e}_t+\mathbf{e}_x\right), & \mathbf{e}_3 &= \mathbf{e}_z,
\end{alignat}
\end{subequations}
where $\gamma=1/\sqrt{1-\beta^2}$ and $\left(\mathbf{e}_t,\mathbf{e}_x,\mathbf{e}_y,\mathbf{e}_z\right)$ are the four base vectors of an observer at rest with respect to the Minkowski coordinate system (\ref{eq:minkowski}). Each base vector points into its positive coordinate direction, where $\mathbf{e}_0$ is adapted to the four-velocity $\mathbf{u}$ of the moving observer,
\begin{equation}
  \mathbf{e}_0 = \frac{1}{c}\mathbf{u}.
\end{equation}
The current four-velocity of the accelerated twin Tina at time $t'$ is given by $\mathbf{u}=u^j\mathbf{e}_j, (j=t,x,y,z)$ with
\begin{equation}
  u^t = c\frac{dt}{dt'} = c\gamma(\beta),\quad u^x=\frac{dx}{dt'}=\beta\gamma(\beta), 
\end{equation}
$u^y=u^z=0$, and her four-acceleration $a^{\mu}$ is
\begin{equation}
 a^t = \frac{d^2t}{dt'^2} = \frac{\alpha}{c}\beta\gamma(\beta),\quad a^x = \frac{d^2x}{dt'^2} = \alpha\gamma(\beta),
\end{equation}
$a^y=a^z=0$, where $\beta=\beta(t')$ is the relative velocity at Tina's proper time $t'$. Thus, her local tetrad at time $t'$ is given by
\begin{subequations}
\begin{alignat}{3}
  \mathbf{e}_0^{S'}(t') &= \cosh\frac{\alpha t'}{c}\mathbf{e}_t+\sinh\frac{\alpha t'}{c}\mathbf{e}_x, &\quad \mathbf{e}_2^{S'}(t') &= \mathbf{e}_y,\\
  \mathbf{e}_1^{S'}(t') &= \sinh\frac{\alpha t'}{c}\mathbf{e}_t+\cosh\frac{\alpha t'}{c}\mathbf{e}_x, & \mathbf{e}_3^{S'}(t') &= \mathbf{e}_z.
\end{alignat}
\end{subequations}
With the help of the local tetrad it is straight forward to derive the aberration and Doppler effect formula. Consider a wave vector $\mathbf{k}$ of an incoming light ray (compare Fig.~\ref{fig:waveVector}).

This wave vector can either be described with respect to Tina's frame $\left\{\mathbf{e}_i^{S'}\right\}_{(i=0,1,2,3)}$ or to Eric's rest frame $\left\{\mathbf{e}_j\right\}_{(j=t,x,z,y)}$. Thus,
\begin{subequations}
\begin{align}
  \label{eq:wv1}
\nonumber  \mathbf{k} &= \omega\left(\mathbf{e}_t - \sin\vartheta\cos\varphi\,\mathbf{e}_x\right.\\
    &\quad\quad - \sin\vartheta\sin\varphi\,\mathbf{e}_y - \cos\vartheta\,\mathbf{e}_z\left.\right)\\[0.5em]
  \label{eq:wv2}
\nonumber   &= \omega'\left(\mathbf{e}_0^{S'} - \sin\vartheta'\cos\varphi'\,\mathbf{e}_1^{S'}\right.\\
     &\quad\quad - \sin\vartheta'\sin\varphi'\,\mathbf{e}_2^{S'}-\cos\vartheta'\,\mathbf{e}_3^{S'}\left.\right)
\end{align}
\end{subequations}
If we transform both representations into coordinates, we can compare each component. Thus, the time component gives the Doppler shift
\begin{equation}
  \label{eq:dopplerShift}
  \omega = \omega'\gamma\left(1-\beta\sin\vartheta'\cos\varphi'\right),
\end{equation}
and from this we get the redshift factor $z$,
\begin{equation}
  \label{eq:redShiftFactor}
  z = \frac{\omega}{\omega'}-1 = \gamma\left(1-\beta\sin\vartheta'\cos\varphi'\right)-1,
\end{equation}
whereas the spatial components give the aberration formulas
\begin{subequations}
  \label{eq:aberration}
\begin{align}
  \cos\vartheta &= \frac{\cos\vartheta'}{\gamma\left(1-\beta\sin\vartheta'\cos\varphi'\right)},\\
  \cos\varphi &= \frac{\sin\vartheta'\cos\varphi'-\beta}{\sin\vartheta\left(1-\beta\sin\vartheta'\cos\varphi'\right)},\\
  \sin\varphi &= \frac{\sin\vartheta'\sin\varphi'}{\gamma\sin\vartheta\left(1-\beta\sin\vartheta'\cos\varphi'\right)}.
\end{align}
\end{subequations}
The inverse formulas for the Doppler shift and the aberration follow from Eqn.(\ref{eq:dopplerShift}) and (\ref{eq:aberration}) by simple interchanging $(\beta,\omega,\vartheta,\varphi)\mapsto (-\beta,\omega',\vartheta',\varphi')$. Since aberration and Doppler shift depend only on the relative directions, we can also use the angle $\chi$ between the wave vector $-\mathbf{k}$ and the direction of motion $\mathbf{e}_x$ with
\begin{equation}
  \cos\chi = \sin\vartheta\cos\varphi.
\end{equation}
From Eq.~(\ref{eq:redShiftFactor}) follows that Tina will see objects to be blueshifted for $z<0$ and redshifted for $z>0$.\par
A similar consideration leads to length contraction. A fixed distance $l'$ to some point in the direction $(\vartheta',\varphi')$ with respect to Tina's current frame would have a length 
\begin{equation}
  \label{eq:lengthContr}
  l=l'\sqrt{\sin^2\vartheta'\left(\gamma^2\cos^2\varphi'+\sin^2\varphi'\right)+\cos^2\vartheta'}
\end{equation}
``measured'' by Eric.
%% ----------------------------------
%%        subsec: apparent size
%% ----------------------------------
\subsection{Apparent size of an object}
\begin{figure}[ht]
  \includegraphics[scale=0.9]{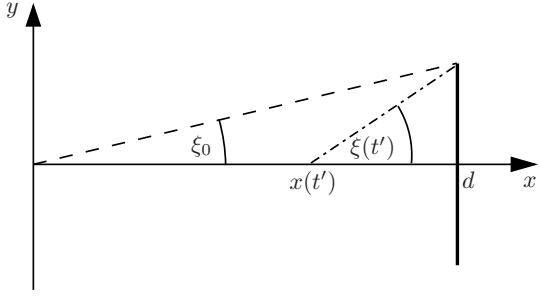}
  \caption{\label{fig:apexAngle}An object, a distance $d$ apart from the origin $x=0$, has an apex angle $\xi_0$. But, an observer at rest at the current position $x(t')$ would measure an apex angle $\xi(t')$.}
\end{figure}

In case of accelerated motion, objects in the direction of motion always seem to recede in the first moments even though one is approaching. The reason for this is also based on the aberration effect. Consider an object a distance $d$ apart with apex angle $\xi_0$ as seen from the origin $x=0$, Fig.~\ref{fig:apexAngle}.

An observer at rest at the current position, Eq.~(\ref{eq:currPos0}), 
\begin{equation}
  \label{eq:accMotion}
  x\left(t'\right) = \frac{c^2}{\alpha}\left(\cosh\frac{\alpha t'}{c}-1\right)
\end{equation}
of the accelerating twin Tina would measure a current apex angle $\xi(t')$ with
\begin{equation}
  \tan\xi(t') = \frac{d\tan\xi_0}{d-x(t')}.
\end{equation}
The acceleration $\alpha$ in Eq.~(\ref{eq:accMotion}) decides whether Tina approaches $(\alpha>0)$ the object or recedes $(\alpha<0)$. From the aberration formula (\ref{eq:aberration}b) follows the apex angle $\xi'(t')$ of the object with respect to Tina,
\begin{equation}
  \label{eq:viewAngle}
  \cos\xi'(t') = \frac{1+\beta(t')\sqrt{1+\tan^2\xi(t')}}{\sqrt{1+\tan^2\xi(t')}+\beta(t')},
\end{equation}
where the velocity $\beta(t')$ is given by Eq.~(\ref{eq:relVel}). Hence, from the partial derivative of Eq.~(\ref{eq:viewAngle}) with respect to Tina's proper time $t'$ follows that
\begin{equation}
  \frac{d\xi'}{dt'}\bigg|_{t'=0} = -\frac{\alpha}{c}\sin\xi_0.
\end{equation}
Thus, accelerating from zero velocity in the direction to the object $(\alpha>0)$ has the effect, that this object appears to shrink in the first moment giving the impression to recede from it instead of approaching. This impression holds until $d\xi'/dt' = 0$,
\begin{equation}
  t'_n = \frac{c}{\alpha}\ln\frac{\psi_1+\sign(\alpha)\delta\sqrt{\left(1+\tan^2\xi_0\right)\psi_2}}{2\left(1+\delta\right)},
\end{equation}
where $\delta=\alpha d/c^2$, $\psi_1 = \left(1+\delta\right)^2+1+\delta^2\tan^2\xi_0$ and $\psi_2 = \left(2+\delta\right)^2+\delta^2\tan^2\xi_0$.\par
An acceleration into the opposite direction results in a magnification which has its maximum at time $t'_n$. Even though a velocity close to the speed of light has a tremendous magnification effect, the distance to the object grows exponentially and predominates the magnification effect.\par
{\it Example:} If Tina leaves home with acceleration $\alpha=-1g\approx-9.81~m/s^2)$ away from the sun, the sun has an apex angle $\xi_0\approx 0.267\degree$ at a distance $d\approx 1.5\cdot 10^{11}~m$. After $t'_n\approx 500~s$ Tina reaches $\beta\approx 1.6\cdot 10^{-5}$ and the maximum magnification is
\begin{equation}
  \frac{\xi'\left(t'_n\right)}{\xi_0}\approx 1-\frac{1}{2}\frac{\alpha d}{c^2}\approx 1+8.2\cdot 10^{-6},
\end{equation}
while the angular size as seen by an observer at rest at her current distance $d(t')\approx d+1.2\cdot 10^6~m$ would be $\xi/\xi_0\approx 1-8.2\cdot 10^{-6}$.
%% ----------------------------------
%%        subsec: apparent position
%% ----------------------------------
\subsection{Apparent position of an object}
\begin{figure}[ht]
  \includegraphics[scale=0.9]{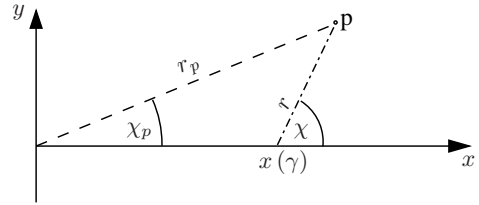}
  \caption{\label{fig:apparentPos}An object $p$ is located at position $(r_p,\chi_p)$ with respect to the initial position $x\left(\gamma=1\right)$ of Tina.}
\end{figure}

{\it What's the apparent position of an object $p$ with respect to the accelerating observer Tina?} If we parameterize Tina's current position, Eq.~(\ref{eq:currPos0}), by her velocity, Eq.~(\ref{eq:relVel}), we get
\begin{equation}
  x = \frac{c^2}{\alpha}\left\{\cosh\left[\artanh(\beta)\right]-1\right\} = \frac{c^2}{\alpha}\left(\gamma-1\right)
\end{equation}
with $\gamma=1/\sqrt{1-\beta^2}\geq 1$. The position $x_p=r_p\cos\chi_p$, $y_p=r_p\sin\chi_p$ of the object $p$, where $r_p>0$ is the distance to the initial position of Tina and $0\leq\chi_p\leq\pi$ is the initial angle with respect to Tina's direction of motion (compare Fig.~\ref{fig:apparentPos}), transforms into Tina's frame according to the aberration formula,
\begin{equation}
  \label{eq:coschi}
  \cos\chi' = \frac{\left(x_p-x\right)/r+\beta}{1+\beta\left(x_p-x\right)/r},
\end{equation}
where the distance $r$ is given by $r^2=\left(x_p-x\right)^2+y_p^2$. 

\begin{figure}[ht]
  \includegraphics[scale=0.7]{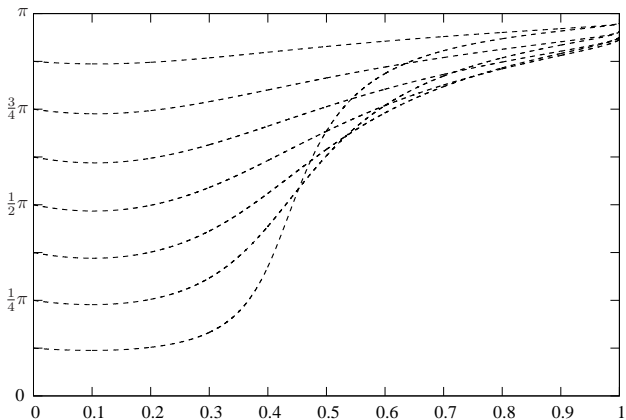}
  \caption{\label{fig:frozen1}The observation angle $\chi'$ is plotted over the velocity $\beta$ for $\mathfrak{A}=c^2/\left(\alpha r_p\right)\approx 9.68$. In the first instance, the objects apparently approaches the center of motion. But for higher velocities, they recede again. Since $\beta=1$ is reached only approximately, an object at $(r_p,\chi_p)$ seems to ``freeze'' at $\chi'_{\text{lim}}=\chi'\left(\beta\rightarrow 1\right)$.}
\end{figure}

\begin{figure}[ht]
  \includegraphics[scale=0.7]{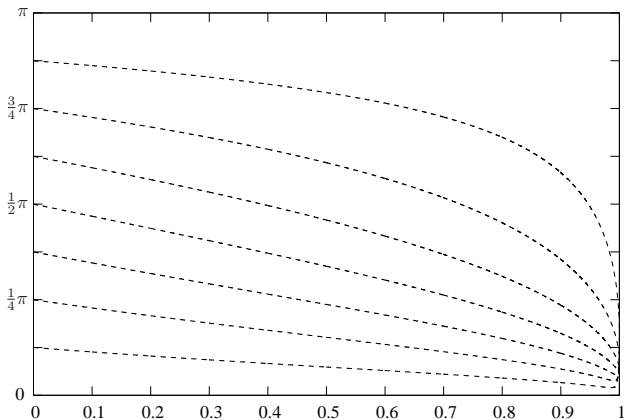}
  \caption{\label{fig:frozen2}The observation angle $\chi'$ is plotted over the velocity $\beta$ for  $\mathfrak{A}=c^2/\left(\alpha r_p\right)\approx 0.0968$. Note that even objects, which are actually behind the observer $(\chi_p>\pi/2)$,  might apparently ``freeze'' in front of the observer.}
\end{figure}

\noindent Eq.~(\ref{eq:coschi}) can be written for $\beta\geq 0$ as
\begin{equation}
  \cos\chi' = \frac{\left[\cos\chi_p-\mathfrak{A}\left(\gamma-1\right)\right]/\tilde{r} + \sqrt{1-1/\gamma^2}}{1+\sqrt{1-1/\gamma^2}\left[\cos\chi_p-\mathfrak{A}\left(\gamma-1\right)\right]/\tilde{r}}
\end{equation}
with abbreviation $\mathfrak{A}=c^2/\left(\alpha r_p\right)$ and 
\begin{equation}
  \tilde{r} = \sqrt{\mathfrak{A}^2\left(\gamma-1\right)^2-2\mathfrak{A}\left(\gamma-1\right)\cos\chi_p+1}.
\end{equation}
The observation angle $\chi'=\chi'\left(\beta\right)$ is shown in Fig.~\ref{fig:frozen1} and Fig.~\ref{fig:frozen2} for two different values of $\mathfrak{A}$.

From the first derivative, $d\chi'/d\gamma=0$, we get an extremum at 
\begin{equation}
  \gamma_e = \frac{1}{2\mathfrak{A}\left(\mathfrak{A}+\cos\chi_p\right)}+1,
\end{equation}
which is only valid if $\mathfrak{A}+\cos\chi_p>0$. The associated velocity $\beta_e$ reads
\begin{equation}
  \beta_e = \frac{\sqrt{1+4\mathfrak{A}\cos\chi_p+4\mathfrak{A}^2}}{1+2\mathfrak{A}\cos\chi_p+2\mathfrak{A}^2}.
\end{equation}
For $\mathfrak{A}+\cos\chi_p>0$ the second derivative follows after some lengthy calculation to be
\begin{equation}
  \frac{d^2\chi'}{d\gamma^2}\bigg|_{\gamma_e} = \frac{\sin\chi_p}{\mathfrak{A}\left(\gamma_e^2-1\right)^2}\geq 0.
\end{equation}
Thus, if there is an extremum, it is always a minimum.\par
In the limit $\beta\rightarrow 1$, $\gamma$ tends to infinity, and an object with coordinates $(r_p,\chi_p)$ seems to ``freeze'' at an observation angle $\chi'_{\text{lim}}$ with respect to Tina's frame,\footnote{The limit can be calculated using de l'Hospital's rule.}
\begin{equation}
  \cos\left(\chi'_{\text{lim}}\right) = \lim\limits_{\gamma\rightarrow\infty}\cos\chi' = \frac{\sin^2\chi_p-\mathfrak{A}^2}{\sin^2\chi_p+\mathfrak{A}^2}.
\end{equation}
Note that two objects $p_1$ and $p_2$ with $r_{p_1}=r_{p_2}$ and $\chi_{p_2}=\pi-\chi_{p_1}$ will ``freeze'' at the same observation angle $\chi'_{\text{lim}}$.\par
As we have seen, the apparent position of an object considerably depends on its position $(r_p,\chi_p)$ and on the acceleration $\alpha$ of the observer. This might be contradictory to one's expectation that an object should appear nearly behind oneself when one is infinitely apart.
% -----------------------------------------------------------------
%           Visualization of the stellar sky
% -----------------------------------------------------------------
\section{\label{sec:visSky}Visualization of the stellar sky}
%% ----------------------------------
%%        subsec: effects
%% ----------------------------------
\subsection{Aberration and Doppler shift}
\begin{figure}[ht]
  \includegraphics[scale=0.7]{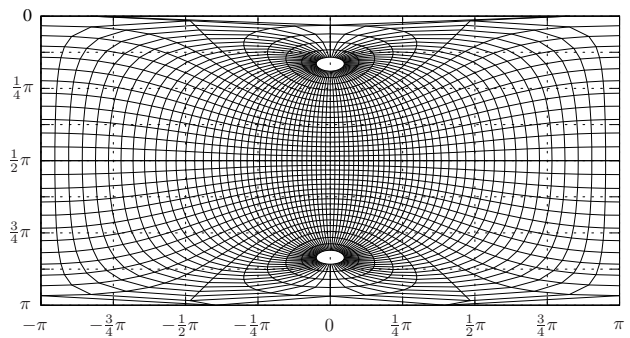}
  \caption{\label{fig:stellarSphereV05}Stellar sky at $\beta=0.5$ in $4\pi$-representation where $\varphi=\left(-\pi,\pi\right)$ is plotted on the abscissa and $\vartheta=\left(0,\pi/2\right)$ is plotted on the ordinate. The center of the image $(\vartheta=\pi/2,\varphi=0)$ corresponds to the direction of motion. The circles of latitude and the meridians are separated by $5\degree\approx 0.087rad$.}
\end{figure}

\begin{figure}[ht]
  \includegraphics[scale=0.7]{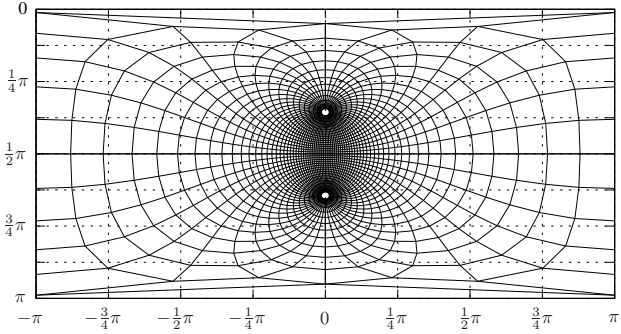}
  \caption{\label{fig:stellarSphereV09}Stellar sky at $\beta=0.9$ in $4\pi$-representation. Because of the aberration the nodes of the stellar sphere move together.}
\end{figure}

{\it What could Tina really see, if she would look out of the windows of her rocket?} If there would be a sphere fixed at infinity with circles of latitude and meridians each separated by $5$ degrees, Tina would see a warped lattice like in Fig.~\ref{fig:stellarSphereV05} and \ref{fig:stellarSphereV09}. In contrast to the visualization of Scott and van Driell\cite{scott1970} we use the $4\pi$-representation where the azimuth angle $\varphi$ and the zenith distance $\vartheta$ are plotted like Cartesian coordinates in order to show the full sky. The disadvantage of this representation is the distortion at the nodes $\vartheta=0$ and $\vartheta=\pi$. Here, the direction of motion corresponds to the center of the representation $(\vartheta=\pi/2,\varphi=0)$.

Because of aberration, Eqn.~(\ref{eq:aberration}a-c), the nodes of the stellar sphere move together according to Tina's velocity. Their angular separation $\Delta\vartheta'$ follows from Eq.~(\ref{eq:aberration}a),
\begin{equation}
  \Delta\vartheta' = \pi-2\arccos\sqrt{1-\beta^2}.
\end{equation}
Since angular distance gets smaller in the direction of motion, an object seems to be farther away in comparison to its real distance. On the other hand, objects in the opposite direction seem to grow.\par
\begin{figure}[ht]
  \includegraphics[scale=0.7]{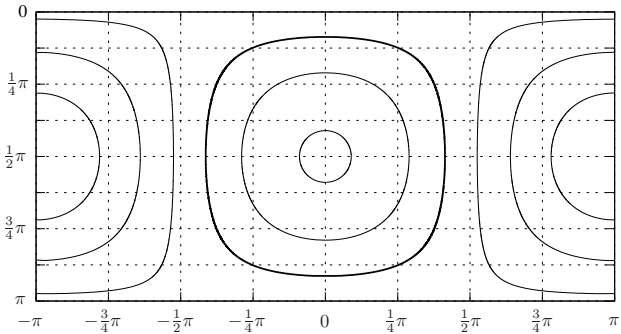}
  \caption{\label{fig:zlines0p5}Lines of constant redshift $z$ at velocity $\beta=0.5$ in $4\pi$-representation. From inside to outside: $z=-0.4$ to $z=0.6$ step $0.2$; the bold line marks $z=0$.}
\end{figure}
\begin{figure}[ht]
  \includegraphics[scale=0.7]{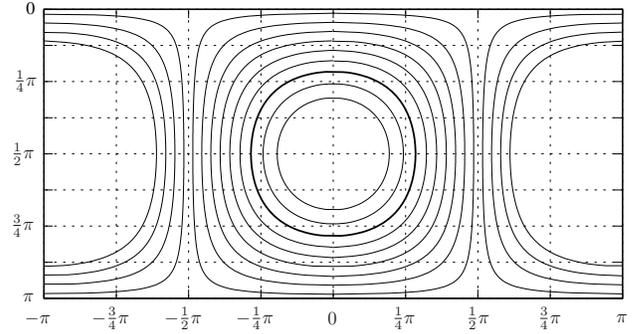}
  \caption{\label{fig:zlines0p9}Lines of constant redshift $z$ at velocity $\beta=0.9$ in $4\pi$-representation. From inside to outside: $z=-0.4$ to $z=2.0$ step $0.2$; the bold line marks $z=0$.}
\end{figure}

Besides the mere geometrical aspects, the following two figures show lines of constant Doppler shift (Fig.~\ref{fig:zlines0p5} and Fig.~\ref{fig:zlines0p9}). As long as the velocity $\beta=0$, there is no Doppler shift, but for velocities $\beta>0$ light is Doppler shifted following Eq.~(\ref{eq:redShiftFactor}),
\begin{equation}
  z = \gamma\left(1-\beta\cos\chi'\right) - 1,
\end{equation}
where $\chi'$ is the angle between the direction of motion and the incoming light ray.

\noindent Zero Doppler shift occurs for $\beta>0$ at an angle $\chi'_0$ with
\begin{equation}
  \label{eq:zeroDoppler}
  \cos\chi'_0 = \frac{1-\sqrt{1-\beta^2}}{\beta}
\end{equation}
and the difference $\Delta z$ between maximum blueshift and maximum redshift equals to
\begin{equation}
  \Delta z = z(\chi'=\pi)-z(\chi'=0) = 2\beta\gamma.
\end{equation}
Thus, the faster Tina will be the more of the sky is redshifted. Only a small portion of the sky in the direction of motion is blueshifted.\par
\begin{figure}[ht]
  \includegraphics[scale=0.7]{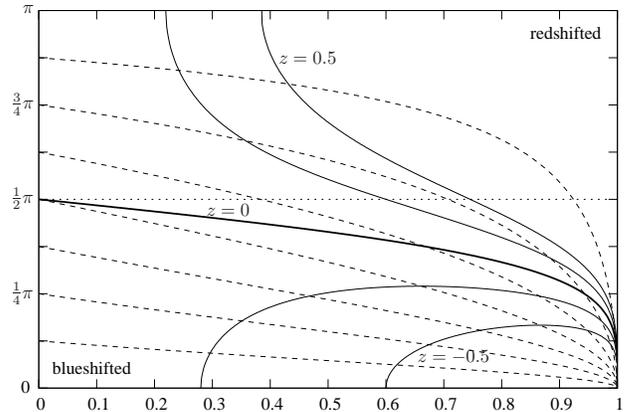}
  \caption{\label{fig:shiftAberr}The observation angle $\chi'$ is plotted over the velocity $\beta$. The solid lines are lines of constant Doppler shift $z$ according to Eq.~(\ref{eq:chipDoppler}), while the dashed lines represent the aberration of an angle $\chi$, compare Eq.~(\ref{eq:chipAberr}).}
\end{figure}

Another interesting detail is shown in Fig.~\ref{fig:shiftAberr}, where the observed angle $\chi'$ is plotted over the velocity $\beta$ according to Doppler shift,
\begin{equation}
  \label{eq:chipDoppler}
  \chi' = \arccos\left(\frac{1-\left(z+1\right)\sqrt{1-\beta^2}}{\beta}\right),
\end{equation}
and aberration
\begin{equation}
  \label{eq:chipAberr}
  \chi' = \arccos\left(\frac{\cos\chi+\beta}{1+\beta\cos\chi}\right).
\end{equation}
For $\beta=0$ we have no Doppler shift and no aberration.

Objects in front of the observer, $\chi<\pi/2$, will always be blueshifted and seem to be in front of him, $\chi'<\pi/2$. But, for objects in the back side of the observer, $\chi>\pi/2$, it depends on the velocity whether they apparently are in front of him or not. Furthermore, an object at an angle $\chi>\pi/2$ will turn from being redshifted to being blueshifted when the velocity $\beta$ will be faster than
\begin{equation}
  \label{eq:betaRedBlue}
  \beta_{\text{red-blue}} = -\frac{2\cos\chi}{1+\cos^2\chi}.
\end{equation}

%% ----------------------------------
%%        subsec: temp and brightness
%% ----------------------------------
\subsection{Temperature and brightness}
For visualizing the stellar sky we use the Hipparcos star catalogue.\footnote{The Hipparcos star catalogue consists of about $118,000$ stars where most of them are at a distance of roughly  $100$pc. The data we are interested in are: (H3) right ascension, (H4) declination, (H5) Magnitude in Johnson V, (H11) trigonometric parallax, (H37) Johnson B-V color, (H71) HD number. The digital catalogue ``I/239'' can be found on {http://cdsweb.u-strasbg.fr/cats/Cats.htx}; 1997HIP...C......0E - European Space Agency SP-1200 (1997) .} Extracting the Johnson 'B-V' color, we assign a temperature $T=T_{B-V}$ to each star by the empirical law\cite{reed}
\begin{equation} 
  \label{eq:BmV}
  B-V = \left\{\begin{array}{cl} C_1\lg(T)+C_2 & \left(\lg T \leq 3.961\right)\\
      C_3\lg(T)^2+C_4\lg(T)+C_5 & \left(\lg T > 3.961\right)\end{array}\right.
\end{equation}
with constants from Tab.~\ref{tab:consts} and the logarithm to the base $10$. This is of course only a limited approximation to the real temperatures of the stars but it simplifies the following calculations.
\begin{table}[ht]
  \setlength{\tabcolsep}{0.25cm}
  \begin{tabular}{cr@{\qquad}|@{\qquad}cr}
    {\bf Coefficient} & {\bf Value} & {\bf Coefficient} & {\bf Value}\\ \hline
    $C_1$ & $-3.684$ & $C_6$ & $-8.499$\\
    $C_2$ & $14.551$ & $C_7$ & $13.421$\\
    $C_3$ & $0.344$  & $C_8$ & $-8.131$\\
    $C_4$ & $-3.402$ & $C_9$ & $-3.901$\\
    $C_5$ & $8.037$  & $C_{10}$ & $-0.438$
  \end{tabular}
  \caption{\label{tab:consts}The coefficients $C_i$ for the Eqns (\ref{eq:BmV}) and (\ref{eq:bc}) are taken from Reed\cite{reed}.}
\end{table}

Furthermore, we need the bolometric correction $(BC)$ in order to transform from the visual $M_V$ to the bolometric magnitude $M_{\text{bol}}$,
\begin{equation}
  M_{\text{bol}} = M_V + BC
\end{equation}
with
\begin{equation}
  \label{eq:bc}
  BC = C_6\mathfrak{t}^4 + C_7\mathfrak{t}^3 + C_8\mathfrak{t}^2 + C_9\mathfrak{t} + C_{10},
\end{equation}
where $\mathfrak{t}=\lg(T)-4$.\par
\begin{figure}[ht]
  \includegraphics[scale=0.55]{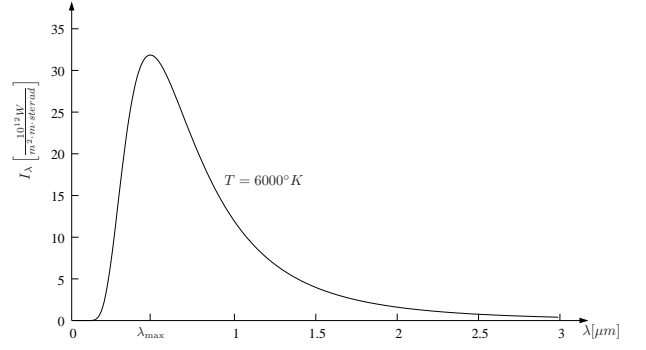}
  \caption{\label{fig:planckSpectr}Planck spectrum at temperature $T=6000\degree K$ with maximum at $\lambda_{\text{max}}\approx 0.483~\mu m$.}
\end{figure}

Instead of the proper spectrum of each star we use a Planck spectrum at temperature $T=T_{B-V}$ with spectral intensity
\begin{equation}
  \label{eq:spectralIntensity}
  I_{\nu} = \frac{2h\nu^3}{c^2}\frac{1}{e^{h\nu/(k_BT)}-1},
\end{equation}
where $h$ is Planck's constant, $c$ is the speed of light and $k_B$ is the Boltzmann constant.\footnote{We take the physical constants from the National Institute of Standards and Technology, http://physics.nist.gov/cuu/Constants.} If the spectral intensity is referred to wave length, we get from $I_{\nu}d\nu=-I_{\lambda}d\lambda$ together with $c=\lambda\nu$ the expression
\begin{equation}
  I_{\lambda} = \frac{2hc^2}{\lambda^5}\frac{1}{e^{hc/(\lambda k_BT)}-1}.
\end{equation}
The typical shape of a Planck spectrum is shown in Fig.~\ref{fig:planckSpectr}, where the wavelength $\lambda_{\text{\tiny max}}$ of the maximum of intensity follows from Wien's displacement law
\begin{equation}
  \label{eq:wienLaw}
  \lambda_{\text{max}} T=b
\end{equation}
with Wien's displacement constant $b=2.8978\cdot 10^{-3}K~m$.

We use a Planck spectrum here, because it simply transforms with the Doppler factor according to the relativistic Liouville theorem \cite{lindquist1966}. From $I_{\nu}/\nu^3=const$ together with Eqn.~(\ref{eq:spectralIntensity}) and (\ref{eq:redShiftFactor}) the temperatures transform like
\begin{equation}
  \label{eq:tempTransf}
  \frac{T_{\text{star}}}{T'_{\text{star}}}=\frac{\nu}{\nu'} = z+1.
\end{equation}
Thus, a blueshifted star with $-1<z<0$ seems to be hotter than it really is compared to its own rest frame.\par
The luminosity $L$ for an isotropically radiating black body is given by\cite{karttunen} $L=4\pi\sigma R^2 T^4$ with radius $R$ of the black body and the Stefan-Boltzmann constant $\sigma\approx 5.67\cdot 10^{-8}W m^{-2}K^{-4}$. Thus, the absolute bolometric magnitude $M_{\text{bol}}$ of an object with a Planck spectrum at temperature $T$ reads
\begin{subequations}
  \begin{align}
    M_{\text{bol}}-M_{\text{bol},\astrosun} &= -2.5\lg\frac{L}{L_{\astrosun}}\\
\label{eq:mbol1} &= -5\lg\frac{R}{R_{\astrosun}}-10\lg\frac{T}{T_{\astrosun}},
  \end{align}
\end{subequations}
where we approximated the spectrum of the sun by a Planck spectrum at temperature $T_{\astrosun}$.\footnote{The absolute bolometric magnitude of the sun is $M_{\text{bol},\astrosun}=4.83$ and the effective temperature is $T_{\astrosun}\approx 5800\degree K$, compare http://nssdc.gsfc.nasa.gov/.} The apparent bolometric magnitude $m_{\text{bol}}$ follows from
\begin{equation}
  m_{\text{bol}} - M_{\text{bol}} = 5\lg\frac{r}{10 pc},
\end{equation}
where $r$ is the distance between star and observer.\par
Now, from Tina's point of view, we have to transform the Planck spectrum of a star into her current rest frame according to Eq.~(\ref{eq:tempTransf}) resulting in a different absolute bolometric magnitude $M'_{\text{bol}}$,
\begin{equation}
  \label{eq:mbol2}
  M'_{\text{bol}} - M_{\text{bol},\astrosun} =  -5\lg\frac{R'}{R_{\astrosun}}-10\lg\frac{T'}{T_{\astrosun}}
\end{equation}
If we make the assumption that the radius $R$ will not be affected by the transformation to Tina's rest frame\footnote{This assumption seems to be reasonable, because a ball keeps its circular outline under Lorentz transformations\cite{hollenbach1976}.}, $R=R'$, the bolometric magnitude $M'_{\text{bol}}$ follows from Eqn. (\ref{eq:mbol1}) and (\ref{eq:mbol2}),
\begin{equation}
  M'_{\text{bol}}-M_{\text{bol}} = 10\lg\left(z+1\right).
\end{equation}
Furthermore, we have to adapt the current distance between Tina and the star. Replacing $r=l$ by $l'$ via Eq.~(\ref{eq:lengthContr}), the apparent magnitude $m'_{\text{bol}}$ as seen by Tina reads
\begin{equation}
  m'_{\text{bol}}-m_{\text{bol}} = 10\lg\left(z+1\right)-5\lg\frac{l}{l'}.
\end{equation}
Since we consider stars as point like sources, we have enough information for visualizing the stellar sky. But if we would come to close to a star we had to take its expansion into account. For more information on how an expanded star would look like we refer the reader to Kraus\cite{kraus2000}.
%% ----------------------------------
%%        subsec: constellations
%% ----------------------------------
\subsection{Constellations}
\begin{figure}[ht]
  \includegraphics[scale=0.72]{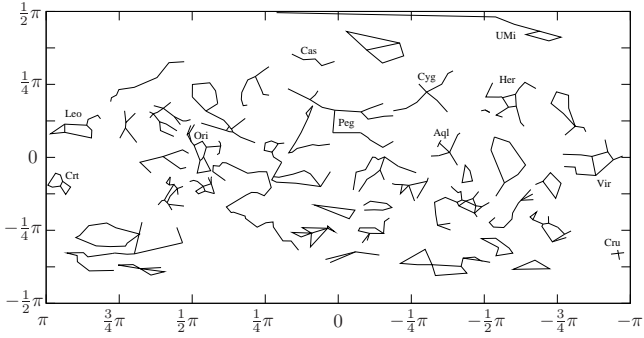}
  \caption{\label{fig:stellarSky0}The stellar sky marked by some constellations as seen at rest. In the $4\pi$-representation the right ascension $\alpha$ is plotted on the abscissa and the declination $\delta$ is plotted on the ordinate. Abbreviations: {\it (Aql) Aquila, (Cas) Cassiopeia, (Crt) Crater, (Cru) Crux, (Cyg) Cygnus, (Her) Hercules, (Leo) Leo, (Ori) Orion, (Peg) Pegasus, (UMi) Ursa Minor}.}
 \end{figure}

\begin{figure}[ht]
  \includegraphics[scale=0.72]{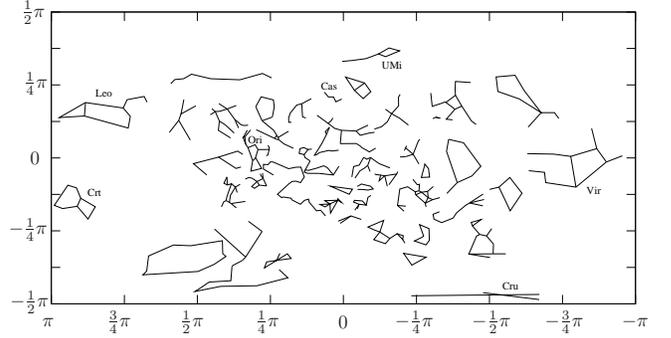}
  \caption{\label{fig:stellarSky50}The stellar sky as seen by an observer passing the Earth with $50$ percent the speed of light. The distortion of the constellation {\it Southern Cross (Cru)} is due to the $4\pi$-projection and the aberration effect, compare Fig.~\ref{fig:stellarSphereV05}.}
 \end{figure}

\begin{figure}[ht]
  \includegraphics[scale=0.72]{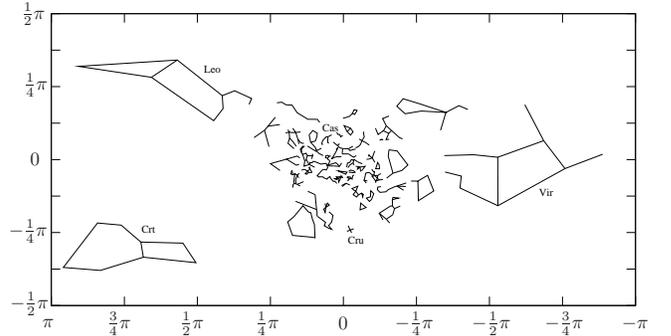}
  \caption{\label{fig:stellarSky90}The stellar sky as seen by an observer passing the Earth with $90$ percent the speed of light.}
 \end{figure}

So far, we presented how aberration and Doppler shift determine the view of the stellar sky as seen by a relativistic observer. The next three figures show the stellar sky of an observer from the position of the Earth but at different velocities in the direction right ascension $\alpha=0$ and declination $\delta=0$, compare Figs.~\ref{fig:stellarSky0}-\ref{fig:stellarSky90}. Note that the declination $\delta$ corresponds to $\pi/2-\vartheta$ in comparison to Fig.~\ref{fig:waveVector}. The stars are connected by lines showing the constellations.\footnote{The constellations were pict from the free planetarium software ``Stellarium''; http://www.stellarium.org.}

As explained in the previous section, the aberration effect lets the constellations apparently shrink in the direction of motion whereas the ones which are behind the observer, like {\it Leo}, {\it Virgo (Vir)} and {\it Crater (Crt)}, seem to grow.

In table \ref{tab:stars} we list the stars building the constellations {\it Orion (Ori)}, {\it Cassiopeia (Cas)} and {\it Southern Cross (Cru)} with their rest frame data. For these three constellations we list the apparent distance $d'$, the temperature $T'$ and the bolometric magnitudes $m_{\text{bol}}$ for the velocities $\beta=0.5$ and $\beta=0.9$ in tables~\ref{tab:starsTransformed} and \ref{tab:starsMagnitude}. The apparent distance $d'$ follows from the inverse form of Eq.~(\ref{eq:lengthContr}) which reads
\begin{equation}
d' = d\left[\gamma^2-\beta^2\frac{1-\cos^2\delta\cos^2\alpha}{\left(1+\beta\cos\delta\cos\alpha\right)^2}\right]^{-1/2},
\end{equation}
where the distance $d$, measured in parsec, is related to the trigonometric parallax $\pi$ via $d=1/\pi$. The redshift factor $z$ is given by the inverse of Eq.~(\ref{eq:dopplerShift}),
\begin{equation}
  z = \frac{\omega}{\omega'}-1 = \frac{1}{\gamma\left(1+\beta\cos\delta\cos\alpha\right)}-1,
\end{equation} 
and determines the temperature $T'=T/(z+1)$.
\begin{table}[ht]
  \setlength{\tabcolsep}{0.15cm}
  \begin{tabular}{ccrrrrr}
    {\bf Abbr.} & $\alpha$ & $\delta\quad$ & $\pi\quad$ & $B-V$ & $T\,\,\,$ & {\bf HIP\,\,}\\ \hline\\[-0.5em]
    $\alpha$ Ori & 1.5497 & 0.1293 & 7.63 & 1.50 & 3488 & 27989\\ %39801\\
    $\beta$ Ori & 1.3724 & -0.1431 & 4.22 & -0.03 & 9077 & 24436\\ %34085\\
    $\gamma$ Ori & 1.4187 & 0.1108 & 13.42 & -0.22 & 19245 & 25336\\ %35468\\
    $\delta$ Ori & 1.4487 & 0.0052 &  3.56 & -0.17 & 15279 & 25930\\ %36486\\
    $\epsilon$ Ori & 1.4670 & -0.0210 & 2.43 & -0.18 & 15903 & 26311\\ %37128\\
    $\zeta$ Ori & 1.4868 & -0.0339 & 3.99 & -0.20 & 17038 & 26727\\ %37742\\
    $\kappa$ Ori & 1.5174 & -0.1688 & 4.52 & -0.17 & 14819 & 27366\\[0.3em] \hline\\[-0.7em] %38771
    $\alpha$ Cas & 0.1767 & 0.9868 & 14.27 & 1.17 & 4287 & 3179\\ %3712\\
    $\beta$ Cas & 0.0400 & 1.0324 & 59.89 & 0.38 & 7025 & 746\\ %432\\
    $\gamma$ Cas & 0.2474 & 1.0597 & 5.32 & -0.05 & 9295 & 4427\\ %5394\\
    $\delta$ Cas & 0.3744 & 1.0513 & 32.81 & 0.16 & 8060 & 6686\\ %8538\\
    $\epsilon$ Cas & 0.4991 & 1.1113 & 7.38 & -0.15 & 13732 & 8886\\[0.3em] \hline\\[-0.7em] %11415
%    $\alpha$ Cru & 3.2577 & -1.1013 & 10.17 & -0.24 & 21259 & 60718\\ %108248\\
%    $\beta$ Cru & 3.3498 & -1.0418 & 9.25 & -0.24 & 20696 & 62434\\ %111123\\
%    $\gamma$ Cru & 3.2776 & -0.9968 & 37.09 & 1.60 & 3277 & 61084\\ %108903\\
%    $\delta$ Cru & 3.2077 & -1.0254 & 8.96 & -0.19 & 16569 & 59747 %106490
    $\alpha$ Cru & -3.0255 & -1.1013 & 10.17 & -0.24 & 21259 & 60718\\ %108248\\
    $\beta$  Cru & -2.9334 & -1.0418 & 9.25 & -0.24 & 20696 & 62434\\ %111123\\
    $\gamma$ Cru & -3.0056 & -0.9968 & 37.09 & 1.60 & 3277 & 61084\\ %108903\\
    $\delta$ Cru & -3.0755 & -1.0254 & 8.96 & -0.19 & 16569 & 59747 %106490
  \end{tabular}
  \caption{\label{tab:stars}Star data of some constellations from Figs.~\ref{fig:stellarSky0},~\ref{fig:stellarSky50} and \ref{fig:stellarSky90}. $\alpha$: right ascension, $\delta$: declination, $\pi$: trigonometric parallax (milliarcsec), $B-V$: Johnson B-V color, $T$: temperature (Kelvin) via Eq.(\ref{eq:BmV}), HIP: Hipparcos number.}
\end{table}

\begin{table}[ht]
  \setlength{\tabcolsep}{0.25cm}
  \begin{tabular}{crrrrr}
    {\bf Abbr.} &  $d_{\beta=0}$ & $d'_{\beta=0.5}$ & $T'_{\beta=0.5}$ & $d'_{\beta=0.9}$ & $T'_{\beta=0.9}$\\ \hline\\[-0.5em]
    $\alpha$ Ori & 131.06 & 125.61 &  4070 &  61.90 &  8153\\
    $\beta$  Ori & 236.97 & 222.56 & 11503 & 109.31 & 24479\\
    $\gamma$ Ori &  74.52 &  70.35 & 23895 &  34.56 & 50135\\
    $\delta$ Ori & 280.90 & 266.08 & 18717 & 130.76 & 38894\\
    $\epsilon$ Ori & 411.52 & 390.64 & 19315 & 192.02 & 39886\\
    $\zeta$ Ori & 250.63 & 238.46 & 20499 & 117.27 & 42038\\
    $\kappa$ Ori & 221.24 & 211.27 & 17562 & 103.98 & 35608\\[0.3em] \hline\\[-0.7em]
    $\alpha$ Cas & 70.08 & 63.34 & 6294 & 31.32 & 14641\\
    $\beta$ Cas & 16.70 & 15.13 & 10190 & 7.48 & 23547\\
    $\gamma$ Cas & 187.97 & 171.11 & 13277 & 84.45 & 30424\\
    $\delta$ Cas & 30.48 & 27.78 & 11457 & 13.71 & 26181\\
    $\epsilon$ Cas & 135.50 & 14.49 & 18943 & 61.30 & 42544\\[0.3em] \hline\\[-0.7em]
    $\alpha$ Cru &98.33 & 98.26 & 19032 & 53.01 & 29046 \\
    $\beta$  Cru & 108.11 & 108.11 & 17998 & 59.70 & 26380 \\
    $\gamma$ Cru & 26.96 & 26.95 & 2766 & 15.31 & 3878 \\
    $\delta$ Cru & 111.61 & 111.60 & 14181 & 62.54 & 20304
  \end{tabular}
  \caption{\label{tab:starsTransformed}The stars of table~\ref{tab:stars} have distance $d'$ (parsec) and temperature $T'$ (Kelvin) at velocities $\beta=0.5$ and $\beta=0.9$ in the direction $\alpha=\delta=0$.}
\end{table}

\begin{table}[ht]
  \setlength{\tabcolsep}{0.25cm}
  \begin{tabular}{cccccc}
    {\bf Abbr.} &  $m_V$ & $BC$ & $m_{\text{bol}}^{\beta=0}$ & $m_{\text{bol}}^{\beta=0.5}$ & $m_{\text{bol}}^{\beta=0.9}$\\ \hline\\[-0.5em]
    $\alpha$ Ori &   0.45 & -2.01 & -1.56 & -2.32 & -6.88\\
    $\beta$  Ori &   0.18 & -0.29 & -0.11 & -1.27 & -6.10\\
    $\gamma$ Ori &   1.64 & -1.95 & -0.31 & -1.38 & -6.14\\
    $\delta$ Ori &   2.25 & -1.36 &  0.89 & -0.11 & -4.83\\
    $\epsilon$ Ori & 1.69 & -1.46 &  0.23 & -0.73 & -5.42\\
    $\zeta$ Ori &    1.74 & -1.63 &  0.11 & -0.81 & -5.47\\
    $\kappa$ Ori &   2.07 & -1.28 &  0.79 & -0.05 & -4.66\\[0.3em] \hline\\[-0.7em]
    $\alpha$ Cas &   2.24 & -0.93 &  1.31 & -0.57 & -5.77\\
    $\beta$ Cas &    2.28 & -0.08 &  2.20 &  0.37 & -4.80\\
    $\gamma$ Cas &   2.15 & -0.32 &  1.83 &  0.07 & -5.06\\
    $\delta$ Cas &   2.66 & -0.16 &  2.50 &  0.78 & -4.35\\
    $\epsilon$ Cas & 3.35 & -1.10 &  2.25 &  0.67 & -4.38\\[0.3em] \hline\\[-0.7em]
    $\alpha$ Cru &   0.77 & -2.21 & -1.44 & -0.97 & -4.14\\
    $\beta$  Cru &   1.25 & -2.14 & -0.89 & -0.29 & -3.24\\
    $\gamma$ Cru &   1.59 & -2.45 & -0.86 & -0.13 & -2.82\\
    $\delta$ Cru &   2.79 & -1.56 &  1.23 &  1.90 & -0.91
  \end{tabular}
  \caption{\label{tab:starsMagnitude}The apparent visual magnitude $m_V$ of the stars of table~\ref{tab:stars} have bolometric magnitudes $m_{\text{bol}}^{\beta}$ at velocities $\beta=0$, $\beta=0.5$ and $\beta=0.9$ in the direction $\alpha=\delta=0$.}
\end{table}

Since {\it Cassiopeia} and {\it Orion} are in front of the observer, $\chi_{\text{Cas}}\approx 1.07$  and $\chi_{\text{Ori}}\approx 1.47$, they will always be blueshifted, compare Fig.~\ref{fig:shiftAberr}. But, the {\it Southern Cross (Cru)}, $\chi_{\text{Cru}}\approx 2.14$, will be redshifted until the observer reaches the velocity $\beta_{\text{red-blue}}\approx 0.83$, which follows from Eq.~(\ref{eq:betaRedBlue}). At $\beta=\beta_{\text{red-blue}}$ Cru has apparently changed already from the back side to the front side of the observer, which follows from Eq.~(\ref{eq:chipAberr}) to happen at
\begin{equation}
  \beta_{\chi'=\pi/2} = -\cos\chi_{\text{Cru}} \approx 0.53.
\end{equation}
We have written an interactive Java applet (App.~\ref{app:javaApplet}) for visualizing the stellar sky at different velocities as seen from the position of the Earth. \jApp

% -----------------------------------------------------------------
%           A trip to the end of the universe
% -----------------------------------------------------------------
\section{A trip to the end of the universe}
As a first step Tina will go on an expedition to the center of our galaxy (SgrA*) $8kpc$ away. As we can read from table \ref{tab:travel}, her maximum speed is only $2.8 ppb$ (parts per billion) beyond the speed of light. 
\begin{table*}[ht]
  \setlength{\tabcolsep}{0.5cm}
  \begin{tabular}{lcccc}
    {\bf Object} & {\bf Distance} & {\bf $\beta_{\text{max}}=v_{\text{max}}/c$} & {\bf Tina's time $2T'$} & {\bf Eric's time $2T$}\\ \hline\\[-0.5em]
    Mars & $0.524~AU\approx 4.4~lm$ & $0.003$ & $\thms{49}{45}{13.4}$ & $\thms{49}{45}{13.7}$\\
    Saturn & $8.53~AU\approx 1.2~lh$ & $0.012$ & $8^{\text{d}}\thms{8}{44}{21}$ & $8^{\text{d}}\thms{8}{44}{38}$\\
    Pluto & $38.81~AU\approx 5.4~lh$ & $0.025$ & $17^{\text{d}}20^{\text{h}}10^{\text{min}}$ & $17^{\text{d}}20^{\text{h}}13^{\text{min}}$\\
    $\alpha$ Cen C (HIP 70890) & $1.29~pc$ & $0.949$ & $\tj{3.54}$ & $\tj{5.84}$\\
    Vega (HIP 91262) & $7.76~pc$ & $0.9975$ & $\tj{6.46}$ & $\tj{27.17}$\\
    SgrA* & $8~kpc$ & $1-2.8\cdot 10^{-9}$ & $\tj{19.74}$ & $26^{\text{ka}}$\\
    LMC & $50~kpc$ & $1-7.1\cdot 10^{-11}$ &$\tj{23.30}$ & $163^{\text{ka}}$\\
    M81 & $2~Mpc$ & $1-4.4\cdot 10^{-14}$ & $\tj{30.45}$ & $6.5^{\text{Ma}}$\\
    END & $13.7\cdot 10^9~ly$ & $1-1.0\cdot 10^{-20}$ & $\tj{45.27}$ & $13.7^{\text{Ga}}$
  \end{tabular}
  \caption{\label{tab:travel}Distance from Earth, maximum speed and proper time of both twins for several stellar destinations. In the solar system we will reach only a few percent of the speed of light. Thus, time dilation can be neglected. However, already in the neighborhood of the solar system time dilation is crucial.}
\end{table*}
    
With this tremendous velocity even the extremely cold microwave background radiation\cite{mather1999} at $T_{\text{cmb}}=2.725\degree$~K comes into the visual regime. But in contrast to one's expectation, the Doppler shifted background radiation would not fill the whole sky. By Wien's displacement law (\ref{eq:wienLaw}) an object must have a temperature of about $T_{780}\approx 3700~K$ to emit its maximum radiation in the red light regime with $\lambda=780~nm$. In order to shift the background temperature to $T_{780}$ the observer has to move with a velocity $\beta_{780}\approx 1-1.07\cdot 10^{-6}$. However, from Eq. (\ref{eq:zeroDoppler}) follows that only in the small region of $\chi'_0<3\degree$ the background radiation is blueshifted. The rest of the sky is redshifted with maximum $z\approx 1767$ at $\chi'=\pi$.\par
On the other hand, the redshift brings us the x-ray and $\gamma$-ray sky down to the visual regime. In the case of x-rays with wavelength of $\lambda\approx 10^{-10}m$ Tina has to fly with $\beta\approx 1-6.6\cdot 10^{-7}$. The much higher velocity $\beta\approx 1-6.6\cdot 10^{-12}$ is needed for $\gamma$-rays with $\lambda\approx 10^{-12}m$.\footnote{For more information on x-rays and $\gamma$-rays we refer the reader to the following missions. X-ray: {\it ROSAT} (http://wave.xray.mpe.mpg.de/rosat, DLR), {\it Chandra} (http://chandra.harvard.edu, NASA), {\it XMM-Newton} (http://sci.esa.int, ESA); $\gamma$-ray: {\it INTEGRAL} (ESA).} Just before the $\gamma$-ray sky, the hydrogen 21cm-line of interstellar gas, resulting from a transition between two hyperfine structure energy levels in the hydrogen atom,  will become visible at a velocity of $\beta=1.0-2.7\cdot 10^{-11}$.\par
In general, a wavelength $\lambda$ will be seen at wavelength $\lambda'$ for a velocity
\begin{equation}
  \beta = \left|\frac{\lambda'^2-\lambda^2}{\lambda'^2+\lambda^2}\right|.
\end{equation}
Thus, the minimum and maximum wavelength at velocity $\beta$ which are transformed into the visual regime follow from
\begin{equation}
  \nonumber \lambda_{\text{min}}=380~nm\sqrt{\frac{1-\beta}{1+\beta}}\quad\text{and}\quad\lambda_{\text{max}}=780~nm\sqrt{\frac{1+\beta}{1-\beta}}.
\end{equation}
%($\rightarrow$missions) http://cxc.harvard.edu, $0.1-10keV$, $E=h\nu=hc/\lambda$.
%XMM-Newton (ESA): http://sci.esa.int, $0.1-15keV$
%$\gamma$-ray: INTEGRAL (ESA), $15keV-10MeV$
%microwave: WMAP: $22-90GHz$, $c=\lambda\nu$. Planck: $25GHz-1THz$

But what happens with the Milky Way itself in the meantime? While Tina's journey to the center of our galaxy lasts only $20$ years with respect to her proper time, the Milky Way ages about $26,000$ years. In that time, our sun with rotational velocity\footnote{The rotational velocity of a star in a galaxy do not follow Newton's law, but can be read from the rotation curve of the galaxy. Confer Brand \& Blitz\cite{brandBlitz1993} for the outer rotation curve of the Milky Way.} $v\approx 220~km/s\approx 7.34\cdot 10^{-4}~ly/a$ will cover a distance of about $19~ly$. In order to fully describe the galactic evolution, we need the position as well as the proper motion of each star in the galaxy. So far, the Hipparcos catalogue consisting of about $118,000$ stars, where most of them are at a distance of about $100~pc$, delivers the best positional star reference $(1~mas)$. From the GAIA\footnote{The aim of the GAIA mission is to collect high-precision astrometric data for the brightest $1$ billion objects, compare http://www.rssd.esa.int/.} mission one expects an accuracy in positional astrometry of about $20~\mu as$. Thus, all stars in our galaxy up to $20$th magnitude should be included.\par 
As announced in the title of this article, Tina could also go on a trip to the ``end of the universe'' within life time. Here, the ``end'' means the maximum distance of about $13.7$ billion light years astronomers are able to observe. For these speculations we neglect the expansion of the universe, which is of course not correct. But it is impossible to know from observations the current status of the universe.\par
When Tina reaches maximum velocity $\beta=1-\epsilon$ with $\epsilon=1.0\cdot 10^{-20}$ in the middle of her journey, the relativistic effects are really extremely dramatic. A $\gamma$-factor of about $7.07\cdot 10^9$ results in a different time rate: $\Delta t'=1s$ corresponds to $\Delta t=224a$! Thus, in roughly $12$ days with respect to Tina's proper time our Sun would have finished one revolution around the center of our galaxy. Zero Doppler shift occurs at an angle $\chi'_0\approx\sqrt[4]{8\epsilon}\approx 3.5as$. The maximum blueshift in the direction of motion is $z\left(\chi'=0\right) \approx\sqrt{\epsilon/2} \approx -1+7.07\cdot 10^{-11}$, whereas the maximum redshift is $z\left(\chi'=\pi\right) \approx \sqrt{2/\epsilon}\approx 1.4\cdot 10^{10}$. Since the transition from redshift to blueshift is quite strong, Tina would see only a very small, bright dot in the direction of motion. On the other hand, most of the sky would be awfully cold and very dark. Navigation by stars would be completely impossible.\par
{\it One might ask if she would be able to see the evolution of the universe?}
%% ------------------------------------- appendix ------------------------------
\appendix
% -----------------------------------------------------------------
%        appendix: Java applets
% -----------------------------------------------------------------
\section{\label{app:javaApplet}Java applets}
The Java applet {\it TwinApplet} handles the diagrams of section \ref{sec:flightVega}. There are three input parameters, two of them are obligatory: the acceleration of the traveling twin Tina in terms of Earth's gravity $g$ and either her travel time or the maximum reachable distance. With these parameters four plots are calculated. They are grouped in two panels: one showing the distance and the velocity of Tina and the other showing the time signals which the twins receive of each other.\par
The acceleration of Tina can be modified between $10^{-6}g$ and $10^3g$. The travel time of one acceleration phase can be entered either into a text field or with a slider. While moving the time slider one can observe the immediate change of the plots and the maximum reachable distance. The maximum time is limited to the time that the twin needs to reach the end of the universe. Alternatively one can choose the travel distance and hence calculate the time needed. It is possible to save either the actual shown plots as PNG-images or the calculated data points in a text file.\par
The Java applet {\it RelSkyApplet} is able to visualize either the stellar sky, the star constellations or the cosmic microwave background, each with different velocities as seen by an observer passing the Earth. The main input parameters for all views are velocity and the line of sight (right ascension and declination). For the stellar sky there are some additional parameters such as the magnitude limiter and the magnitude scale. For the stellar sky and the star constellations it is possible to increase the velocity stepwise (without changing position) to observe the rise of effects. By varying the velocity, the user gets a better understanding of how relativistic aberration and Doppler shift affects the visualization.\par
It is possible to choose between two views for the visualizations: A $4\pi$-representation or an hemispherical representation. In both cases, the line of sight corresponds to the center of the representation and can be set manually. The stellar sky can also be seen from any desired point in our galaxy (either in equatorial coordinates or in galactic coordinates). By moving the mouse over the sky-view, the sky coordinates of the actual mouse position are displayed in the lower left corner of the applet. As an additional assistance, longitude and latitude of the seen sky can be displayed as a grid, which deforms according to the increasing velocity.\par
The stars are realized as small disks whose colors are calculated according to appendix \ref{app:spectromToColor} with the star temperatures. It is possible to change the color mode in order to examine a special temperature range. In manual mode the displayed colors correspond merely an arbitrary color scale. The size of the disk represent the star's apparent bolometric magnitude $m_{\text{bol}}$.

By clicking on a star its Hipparcos identifier (HIP) and (if available) its common name are displayed. Another click on "Get Info" provides further information about the star: the complete Hipparcos data set, as well as some of its ``real'' values compared to its ``apparent'' values.

The star constellations view is suited for observing the distortion due to the aberration effect. Bear in mind that the distortion of some constellations in the border area of the $4\pi$-representation might be caused by the representation itself.\par
The basic data of the cosmic microwave background (CMB) was obtained from WMAP.\footnote{Data is taken from http://map.gsfc.nasa.gov} In order to speed up the applets execution, the original data was downgraded to a $4\pi$-representation with a solution of $720\times 360$ pixels. By selecting the microwave mode the color scale is automatically set to a temperature range of $2.5$ to $2.9$ K (which corresponds to the microwave wavelengths). However, due to the Doppler effect, it is also possible to observe the microwave background even in the visual regime (visual color mode) starting at a velocity of about $0.999c$.

% -----------------------------------------------------------------
%        appendix: numerical calculations
% -----------------------------------------------------------------
\section{Numerical calculations}
As we can read from table~\ref{tab:travel}, we have to deal with velocities $\beta$ which come quite close to the speed of light. Since computers can handle only a limited number of digits, the velocity $\beta=1-10^{-5}$ will be stored as single-precision floating-point number as
\begin{align*}
  \hat\beta &= 1.11111111111111101011000_2\times 2^{-1}\\
  & \approx 0.9999899864_{10},
\end{align*}
where $\hat\beta$ is expressed either in normalized binary form (subscript $2$) or in decimal form (subscript $10$).\footnote{Single-precision floating-point numbers are stored in a 32-bit word, whereas double-precision ones are stored in a 64-bit word. See {\it The IEEE-754 Standard for Binary Floating-Point Arithmetic, 1985} for further details.} Of course, this machine impreciseness propagates. Hence, a straight forward numerical calculation of the $\gamma$-factor results in
\begin{align*}
  \hat\gamma &= 1.10111110111010010000011_2\times 2^7\\
  & \approx 223.455123_{10}
\end{align*}
On the other hand, the series expansion of the $\gamma$-factor would read
\begin{equation}
  \gamma=\frac{1}{\sqrt{1-\left(1-\varepsilon\right)^2}} \approx \frac{1}{\sqrt{2\varepsilon}} +  {\cal O}\left(\varepsilon^{1/2}\right)
\end{equation}
and the evaluation for $\varepsilon=10^{-5}$ gives
\begin{align*}
  \hat\gamma &= 1.10111111001101101010111_2\times 2^7\\
  &\approx 223.606796_{10}.
\end{align*}
From the relative errors $E^{\text{rel}}$ for the straight forward calculation and the series expansion,
\begin{equation*}
  E_{\text{straight}}^{\text{rel}} \approx 6.8\cdot 10^{-4}\quad\text{and}\quad E_{\text{series}}^{\text{rel}} \approx 2.5\cdot 10^{-6},
\end{equation*}
we conclude, that it's important to replace an equation by its series expansion for high velocities.

% -----------------------------------------------------------------
%        appendix: From spectrum to color
% -----------------------------------------------------------------
\section{\label{app:spectromToColor}From spectrum to color}
The human visual perception of wavelengths lies in the range between $380nm$ and $780nm$. There are three types of cones on the retina which are sensible to almost red, green and blue light. Hence, any visible color can be composed out of three primary colors.\footnote{This fact is known as the tristimulus theory.}\par
\begin{figure}[ht]
  \begin{center}
    \includegraphics[scale=0.72]{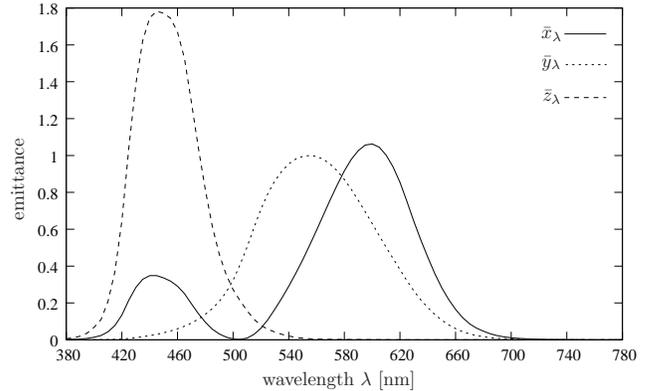}
    \caption{\label{fig:cieXYZ}CIE 1931 2-deg color matching functions $\bar{x}_{\lambda},\bar{y}_{\lambda},\bar{z}_{\lambda}$.}
  \end{center}
\end{figure}

In 1931, the {\it Commission Internationale de l'{\'E}clairage (CIE)} defined three artificial primary colors $\mathbf{X}$, $\mathbf{Y}$ and $\mathbf{Z}$ with the corresponding color matching functions\footnote{Color matching functions can be found at the Institute of Ophthalmology, http://cvrl.ioo.ucl.ac.uk/basicindex.htm.} $\bar{x}(\lambda)$, $\bar{y}(\lambda)$ and $\bar{z}(\lambda)$, compare Fig.~\ref{fig:cieXYZ}.

Any color $\mathbf{C}=X\mathbf{X}+Y\mathbf{Y}+Z\mathbf{Z}$ can be composed out of these primary colors, where the components $X$, $Y$ and $Z$ follow from the spectral intensity distribution $I(\lambda)$ by convolution with the color matching functions, e.g.,
\begin{equation}
  X = k\int I\left(\lambda\right)\bar{x}(\lambda)d\lambda.
\end{equation}
Since we are only interested in the chromaticity values $x$, $y$ and $z$ with $x=X/(X+Y+Z)$ and $y=Y/(X+Y+Z)$ and $z=1-x-y$, the constant $k$ vanishes. In order to get the $rgb$ values for a specific device, we need its primary chromaticity values $x_{(r,g,b,w)}$ and $y_{(r,g,b,w)}$ for red, green, blue and the white point.\footnote{We use the color rendering of spectra by John Walker (http://www.fourmilab.ch).} After the transformation from $xyz$ to $rgb$, we normalize the $rgb$ values to the maximum of them,
\begin{equation}
  (r,g,b) \mapsto\frac{(r,g,b)}{\text{max}(r,g,b)},
\end{equation}
which has the effect, that all colors have their largest possible luminance.

% -----------------------------------------------------------------
%                          Acknowledgments   
% -----------------------------------------------------------------
\acknowledgments
The authors would like to thank Prof. Hanns Ruder for the idea to this work and Prof. J{\"o}rg Frauendiener for many discussions and for carefully reading the manuscript.
This work is supported by the Deutsche Forschungsgesellschaft (DFG), SFB 382, Teilprojekt D4.

\bibliography{lit_twin}% Produces the bibliography via BibTeX.
\end{document}